# Manipulate the Electronic State of Mott Iridate Superlattice through Protonation Induced Electron-Filling

*Meng Wang, Lin Hao, Fang Yin, Xin Yang, Shengchun Shen, Nianlong Zou, Hui Cao, Junyi Yang, Nianpeng Lu, Yongshun Wu, Jianbing Zhang, Hua Zhou, Jia Li\*, Jian Liu\*, and Pu Yu\**

Dr. M. Wang, Dr. S. Shen, N. Zou, Y. Wu, J. Zhang, Prof. P. Yu
State Key Laboratory of Low Dimensional Quantum Physics and Department of Physics, Tsinghua University, Beijing 100084, China
Email: yupu@tsinghua.edu.cn

Dr. L. Hao, J. Yang, Prof. J. Liu
Department of Physics and Astronomy, University of Tennessee, Knoxville, TN 37996, USA.
Email: jianliu@utk.edu

F. Yin, Dr. X. Yang, Prof. J. Li
Guangdong Provincial Key Laboratory of Thermal Management Engineering and Materials and Institute of Materials Research, Tsinghua Shenzhen International Graduate School, Tsinghua University, Shenzhen 518055, China.
Email: li.jia@sz.tsinghua.edu.cn

Dr. H. Cao, Prof. H. Zhou
Advanced Photon Source, Argonne National Laboratory, Lemont, IL 60439, USA

Prof. N. Lu
Beijing National Laboratory for Condensed Matter Physics, Institute of Physics, Chinese Academy of Science, Beijing 100190, China

Prof. P. Yu
RIKEN Center for Emergent Matter Science (CEMS), Wako 351-198, Japan

Prof. P. Yu
Frontier Science Center for Quantum Information, Beijing 100084, China

**Abstract**: Spin-orbit-coupled Mott iridates show great similarity with parent compounds of superconducting cuprates, attracting extensive research interests especially for their electron-doped states. However, previous experiments are largely limited within a small doping range due to the absence of effective dopants, and therefore the electron-doped phase diagram remains elusive. Here we utilize an ionic-liquid-gating induced protonation method to achieve electron-doping into a 5d

Mott-insulator built with SrIrO₃/SrTiO₃ superlattice, and achieve a systematic mapping of its electron-doped phase diagram with the evolution of the iridium valence state from 4+ to 3+, equivalent to doping of one electron per iridium ion. Along increasing doping level, the parent Mott-insulator is first turned into a localized metallic state with gradually suppressed magnetic ordering, and then further evolved into a nonmagnetic band insulating state. This work forms an important step forward for the study of electron-doped Mott iridate systems, and the strategy of manipulating the band filling in an artificially designed superlattice structure can be readily extended into other systems with more exotic states to explore.

## 1. Introduction

$5d$ transition metal oxides (TMO) are attracting increasing research interests, where a plethora of exotic phenomena emerges due to the involved strong spin-orbit coupling (SOC)[1-6]. Generally, $5d$ oxides show wider bandwidth ($W$) with weaker electron correlation, as compared with localized $3d$ systems, hence it is easier to form metallic ground state. However, the dimensionality effect in square-lattice iridates (e. g. $Sr_2IrO_4$ and $SrIrO_3/SrTiO_3$ superlattice) leads to an exotic antiferromagnetic (AFM) Mott insulating state [3-5]. In $IrO_6$ octahedra, the strong SOC splits the degenerated $t_{2g}$ into a full-filled $J_{eff} = 3/2$ and a half-filled $J_{eff} = 1/2$ state (**Figure 1a**). With the insertion of insulating blocks into the $IrO_6$ layers, the suppressed interlayer hopping narrows the bandwidth $W$, and finally opens a Mott gap as the $W$ is smaller than the on-site Coulomb interaction $U$ in the monolayer limit [3-8] (as shown in **Figure 1b, c**).

Inspired by the fact that the resulted $J_{eff} = 1/2$ AFM Mott state has a great similarity with the $S = 1/2$ AFM Mott state in the parent compounds of High-$T_c$ cuprates, theoretical models building on a 2D Ir-O layer predict that a novel superconducting state may emerge from square-lattice iridates upon suitable electron-doping [9-11]. $Sr_2IrO_4$ is the most extensively studied square lattice iridate in the last decades on this route, in which several strategies have been explored to perform electron-doping, e.g. deposition of alkali-metals at sample surface [12-14], substitution of $Sr^{2+}$ with $La^{3+}$ [15-22], and formation of oxygen vacancies through annealing [21,23]. However, hindered by the instability of reduced iridates [24] as well as low solid solubility, all these experiments are limited to either local doping at surface or small doping-level (see Supplementary Table 1) [12-23]. It remains unexplored for highly electron-doped Mott iridates and a completed phase diagram with electron-doping is still missing. Hence, from material science perspective, an important question naturally emerges as how we can realize an effective and pronounced electron-doping into the Mott iridates and how its electronic state will respond to such modulation.

Recently, electric-field controlled proton intercalation (**Figure 2a**, inset) emerges as a new pathway to realize continuous electron-doping in material systems. In this method, the proton, as the smallest ion in nature, is generated from the electrolyzed $H_2O$ within ionic liquid, which can be intercalated into samples with the application of electric-field [25]. Due to the principle of charge neutrality, the positively charged proton would be associated with one electron to complete the structural transformation as protonation (**Figure 1b, c**), which can lead to the modulation of pronounced bulk electron doping with emergence of exotic electronic states within a series of complex oxide systems [25-32] and even in iron-based superconductors [33]. We note that such approach forms an important strategy to manipulate electron doping to complement the well-established chemical substation, oxygen vacancy formation as well as alkali metal intercalation in complex oxide systems.

Here we successfully extend the ILG induced protonation method into an iridate system to map its exotic electron-doped phase diagram. It is important to note that the previous ionic liquid gating (ILG) studies on $Sr_2IrO_4$ shows only negligible modulation effect [22], and the sample is easily damaged during ILG and therefore not appropriate for our proposed protonation study. Instead, we find that the $SrTiO_3/SrIrO_3$ superlattice (SL), which also has the 2D Ir-O layer as the key building block but with a much simpler crystalline and magnetic structure as compared with $Sr_2IrO_4$ [7-9,34,35], and more importantly the SL structure is very robust against the ILG as revealed recently in LSMO/SIO superlattice [30], which makes this study practically. In the SL, the insulating $SrTiO_3$ layer forms a robust crystalline framework to stabilize the quasi-two-dimensional nature of Ir-O layers [34], in which the $J_{eff} = 1/2$ pseudo-spins are antiferromagnetically ordered and have the same translational symmetry along the c-axis as C-type AFM structure [7,8]. Furthermore, the mono-layered iridate features a large in-plane $IrO_6$ octahedral rotation, which generates a finite spin-canting induced net moment via Dzyaloshinskii-Moriya (DM) interaction [8] (Supplementary Figure 1). Importantly, previous theoretical calculation revealed that the Ti-$t_{2g}$ orbital in $SrTiO_3$ locates at a much higher energy than the upper Hubbard band (UHB) [7] as shown in **Figure 1c**, and therefore the electron doping will have negligible influence from the band structure of SrTiO$_3$ layer. With these facts, this SrTiO$_3$/SrIrO$_3$ SL hosts all the crucial elements for exploring the electron doping effect in Mott iridate.

In this work, utilizing the electric-field controlled proton intercalation (protonation process), we achieve a dramatic manipulation of the Hubbard band electron-filling in SrTiO$_3$/SrIrO$_3$ SL, with iridium ions reduced continuously from +4 to near +3. With such electron modulation, we observe the emergence of a novel electronic state with metallicity at high temperature (> 170 K) and a weak insulating behavior at lower temperature region; and then with further increase of the doping level, the sample turns into a band insulator. The magnetoresistivity (*MR*) measurements suggest a suppressed magnetic ordering at low-temperature region in the doped sample until the doping level reaches the full-filled band insulator with a paramagnetic state. Our first-principles calculations reveal that during the protonation process, the protons form a direct bonding with the equatorial oxygen ions at Ir-O layers, leading to an enhanced octahedral rotation, which then competes with the electron doping to play a key role to determine the corresponding electronic and magnetic states in such system.

## 2. Results and Discussion

High-quality SrIrO$_3$/SrTiO$_3$ SLs were grown with pulsed laser deposition (see Experimental Section). **Figure 1d** shows the X-ray diffraction (XRD) 2$\theta$-$\omega$ scan of a representative SL grown on SrTiO$_3$ (001) single crystal substrate, where the sharp diffraction peaks with nice thickness fringes confirm the alternating stacking of SrIrO$_3$ and SrTiO$_3$ unit cells along c-axis with atomic-scale precision. The reciprocal space mapping (RSM) result (**Figure 1e**) suggests that the SL is fully strained with the substrate. The temperature dependent sheet resistance (*RT*), shown in Supplementary Figure 2, presents a typical insulating behavior in this SL with an anomalous kink feature at ~135 K related to the AFM transition [5,6].

To explore the possibility of ILG induced protonation, we first carried out *in-situ* XRD measurements around the SL (004) diffraction peak during ILG to monitor the corresponding structural transformation. With the increase of gating voltage, three distinct regimes can be identified, as revealed in **Figure 2a**. When the gate-voltage ($V_G$) is smaller than 1.0 V, both position and width of the XRD peak remain almost unchanged, which suggests that the proton intercalation effect is negligible. Above this threshold voltage, the (004) peak shifts gradually toward lower angle, and eventually saturates at $V_G$ ~ 2.5 V. It is important to note that the SL remains fully strained through the ILG process (Supplementary Figure 3) without any notable sample degradation. We summarize the structural transformation during ILG in **Figure 2b**, where the c-axis lattice parameter increases from 7.921 Å at $V_G$ = 0 V to 8.154 Å at $V_G$ = 2.5 V, with the lattice expansion of ~3%. We notice that such a structural expansion is consistent with the previous study of proton intercalated oxides [25,29] as well as lightly doped Sr$_2$IrO$_4$ systems [16,21], pointing to the same origin as of electron doping. To further verify the proton intercalation process during ILG, a secondary ion mass spectrometry (SIMS) measurement was carried out on a typical gated sample. Although the phase transformation is volatile (Supplementary Figure 4), we found distinct hydrogen concentration within the gated sample as comparing with the pristine one (Supplementary Figure 5). This fact is further supported by our first-principles calculations, which show that the proton intercalation process has much reduced formation energy as comparing with the formation of oxygen vacancy (Supplementary Figure 6 and Table 2). More specifically, the hydrogen ions (protons) prefer to form bonding with the equatorial oxygen of Ir-O layers as comparing with other sites (apical oxygens or equatorial oxygens of Ti-O layers). We also calculated the lattice constant of proton intercalated SLs with different hydrogen contents, which is quantitively consistent with our experimental results (**Figure 2b**).

With the confirmed structural transformation through ILG, we then carried out *in-situ* X-ray absorption spectroscopy (XAS) measurements with a glancing incidence geometry at both Ir *L-edge* and Ti *K-edge* to probe the corresponding valence state evolution of the SL (see Experimental Section). As shown in **Figure 2c**, we observed a striking red shift (~ 0.8 eV) of the Ir $L_3$-white line in the gated SL ($V_G$ = 2.5 V) as compared with the pristine one. In iridate systems, the peak shift was previously evidenced to be roughly linear-dependent on the electron filling in $t_{2g}$ band, with the red shift of 0.6 eV ~ 0.9 eV through the reduction of $Ir^{4+}$ to $Ir^{3+}$[36, 37]. Therefore, we can deduce that the pristine $Ir^{4+}$ ions were almost completely reduced to $Ir^{3+}$ through ILG in the current study. On the other hand, there is no obvious change on either peak position or profile for Ti *K*-edge (**Figure 2d**), suggesting there is no observable change on its valence state and SrTiO$_3$ unit cell remains intact upon protonation. This should be attributed to the much higher threshold voltage of ILG for SrTiO$_3$ layer [38]. The XAS result thus confirms that the ILG induced protonation in this SL leads to a pronounced electron-filling effect into the $J_{eff}$ = 1/2 band.

Our subsequent *in-situ* electronic transport measurements reveal an interesting evolution of electronic state upon the ILG induced protonation. As summarized in **Figure 3a** and **3b**, a step-by-step increase of $V_G$ from 1.1 V leads to gradually reduced electronic resistivity of the SL and eventually triggers an emergent metallic state at high-temperature region (see details in Supplementary Figure 7a and 7b). The further increase of $V_G$ expands gradually the metallic regime and eventually reaches the lowest sheet resistance at $V_G$ = 1.5 V. However, further increase of $V_G$ quickly destroys the metallic state with clearly enhanced resistance and eventually the insulating state is formed with $V_G$ = 2.3 V. A direct comparison reveals that the resistance of the newly formed insulating state is about one order of magnitude larger than the pristine one, indicating the formation of a new bounded electronic state. To trace the evolution of carrier concentration with ILG, we performed Hall measurements at two typical temperatures of 150 K and 120 K. As shown in Supplementary Figure 7c, the negative slope of all curves suggests the electronic nature of the carriers, which is consistent with the scenario of protonation induced electron-doping. We performed a fitting with single band model to estimate the carrier density (e.g. -1/e$R_H$) at different states (**Figure 3c**). The main trend of the carrier density evolution is consistent with the resistance measurements, with the largest charge density obtained around the metallic state (1.5 - 2.0 V). When changing the temperature from 150 K to 120 K, the carrier density is clearly suppressed, which is consistent with the insulating transport behaviors at low temperatures as well.

In addition to the evolution of electronic states, we also observed systematic change of magnetic states upon protonation, in which both the long-range antiferromagnetic ordering temperature ($T_N$) and the longitudinal spin fluctuation are suppressed. Although the AFM insulating ground state can be captured by a Hubbard model similar to the AFM Mott insulator $La_2CuO_4$, its smaller $U/t$-value for $J_{eff} = 1/2$ electrons leads to a much smaller Mott gap, allowing significant charge fluctuations and hence longitudinal spin fluctuations [39]. The previous studies reveal that external magnetic field can be employed to suppress such spin-charge fluctuations due to the hidden SU (2) symmetry of the $J_{eff} = 1/2$ square lattice with octahedral rotation, giving rise to an anomalous positive magnetoresistance (MR) that is maximized around $T_N$ [39]. Indeed, the pristine SL shows the largest in-plane *MR* at 9 T around 135 K, as shown in **Figure 3d** and Supplementary Figure 2b, and this anomalous temperature-dependence mimics the longitudinal spin susceptibility and affords a convenient route to monitor the AFM transition and the spin fluctuations. During the ILG, with increasing $V_G$ above threshold voltage, the peak of *MR* shifts systematically toward lower temperatures, implying that the AFM order is gradually suppressed along protonation. This downshift is particularly significant when $V_G$ increases from 1.3 to 1.5 V, indicating a collapse of the Mott state and consistent with the emergence of metallicity. Meanwhile, the amplitude of MR is also suppressed from ~6% at 1 V to

~1% at 1.5 V. If one proportionates the AFM order parameter and $T_N$ at the mean field level, this voltage-dependence indicates that the longitudinal spin fluctuations are suppressed first before the reduction of the ground-state staggered magnetization. It becomes difficult to monitor $T_N$ through MR when increasing $V_G$ beyond 1.5 V, because the critical behavior borne by the anomalous temperature-dependence is highly suppressed. Interestingly, the magnetic hysteresis of the MR persists and becomes negative at low temperatures with further increasing $V_G$ till 2.0 V (**Figure 3d** and Supplementary Figure 8), which might suggest the emergence of a weak ferromagnetic state during ILG. When $V_G$ reaches 2.3 V, the *MR* is dramatically suppressed over the whole temperature range without any significant temperature-dependence, which suggests that the nearly filled upper Hubbard band (with all-filled low spin $5d^6$ state) totally destroys the AFM state to form a paramagnetic (PM) state.

To obtain deeper physical insights of the low-temperature insulating state at the optimal doped samples, we performed a systematic analysis of the temperature dependent sheet resistance, at shown in **Figure 3e**. It is interesting to note that the *RT* curves can be fitted nicely with a $R$-$T^{-1/3}$ dependence at a wide temperature region, which indicates the states near the Fermi level are weakly localized [40]. As a comparison, the two-dimensional variable range hopping model (Supplementary Figure 7d) [41] fails to provide a reasonable fitting to these data. Considering the fact that this large temperature range is far above the magnetic transitions, we speculate that apart from the spin fluctuation, some short-range charge ordering or charge glass-like states [40,42] may contribute to the localization.

Combining knowledge of both electronic and magnetic evolutions, we tentatively build up an electron-doped phase diagram of Mott iridate SL as shown in **Figure 3f**. The parent phase as well as the ILG state with small voltage (less than 1.4 V) remains a spin-orbit coupled Mott insulator with the AFM transition temperature being slightly reduced after ILG. With further increase of carrier concentration close to optimal doping through ILG (1.45 V < $V_G$ < 1.7 V), a PM metallic region emerges at high temperature regions. However, at lower temperatures, the ILG samples demonstrate a surprising localized weak-insulating (LWI) behavior with dramatically suppressed long-range AFM state and possibly emergent weak ferromagnetic state (WFM). With further increase of the ILG voltage (i.e. 1.7 V < $V_G$ < 2.0 V), the samples reach an over-doped range with resistance increasing toward insulating state, where the magnetism shows a crossover from WFM to PM state. Finally, in the fully gated samples ($V_G$ > 2.3 V), the UHB is completely filled and the sample turns into a $5d^6$ low-spin PM band insulator.

To understand the evolution of electronic states during the protonation, we carried out first-principles calculations for different proton intercalated $SrTiO_3/SrIrO_3$ SL (see Experimental Section). The structural analysis studies reveal that the proton prefers to bond directly with the equatorial oxygen ion in Ir-O planes (**Figure 4a** and Supplementary Figure 6). With this constructed crystalline structure, we then performed corresponding electronic structure calculations, as summarized in **Figure 4b-e**. The pristine superlattice is a typical Mott insulator with $J_{eff}$ = 1/2 band splitting into an upper Hubbard band (UHB) and a lower Hubbard band (LHB), where the LHB intertwined with the $J_{eff} = 3/2$ orbits (Fig.4b), which is consistent with previous report [7]. As increasing the doping level to 0.25 (Fig.4c), the degenerated UHB further splits into an occupied band below Fermi energy and three unoccupied bands above Fermi energy with the formation of a very small band gap, and in this case the overlapping between the occupied 1/2 and 3/2 orbits is enhanced. Further increasing the doping level to 0.5 (Fig.4d), the pristine UHB is half filled but with a tiny density of states at the Fermi energy, indicating a bad metal phase. As increasing the doping level to 1 (Fig.4e), both 1/2 and 3/2 orbits are intertwined and filled, and therefore the material turns into a band insulator. The trend of such band structure evolution is consistent with the results of our transport experiments, in which a metallic state is observed only at intermediate states, while both the pristine and fully gated samples are insulating. Besides, the calculated lattice structures reveal that the octahedral rotation angle increases gradually along with the proton intercalation, as shown in **Figure 4f**. Considering that the reduction of iridium ion from +4 to +3 tends to dramatically enlarge the radius of Ir-ion as well as the Ir-O bond length, while the in-plane lattice is confined within the substrate, which will lead to the enhancement of the octahedral $c^-$ rotation upon protonation [7,8]. Additionally, the calculations of 0.25 and 0.5 proton doped phases suggest the formation of weak ferromagnetic states (Supplementary Figure 9), which is consistent with the obtained enhanced negative magnetoresistance (at 2K) for the intermediate states (Supplementary Figure 8f).

Prior experiments with light oxygen-vacancies and $La^{3+}$ dopants all indicate a residual magnetic order at low temperature even though the doped iridates have been very close to metallic states [19,23]. As a comparison, our results suggest that the magnetic order in Mott iridate is rather robust before the UHB completely filled. Furthermore, the protonation induced electron-doping also shows a localized weak-insulating ($R$-$T^{-1/3}$ dependence) behavior, which have not been achieved before in iridate systems. To account for the observed exotic state, we consider two possible scenarios. The first one is related with the enhanced scattering effect through random distributed hydrogen ions (protons) within lattice. However, according to our previous study in $5d$ $WO_3$ [26], the disordered hydrogen ions tend to introduce Andersen localization [43] with only weak insulating characters observed at much reduced temperatures (< 20 K), which is in sharp contrast with the notable resistive state with large temperature region at current system. Besides, previous studies reveal that defects in strong correlated two-dimensional cuprates tends to induce resistive state through Mott variable range hopping (VRH) mechanism [44], rather than the weaker localization model. Another mechanism is related with the enhanced $IrO_6$ octahedral rotation. We speculate that the enhanced octahedral rotation hinders but is not enough to totally suppress the hopping of the doped $5d$ electrons [11,44]. Furthermore, the octahedral rotation pattern may also drive the current system away from its counterpart cuprates [11,45], for example the hole doped Mott nickelate $La_{2-x}Sr_xNiO_4$ system shows a robust insulating ground state throughout whole doping range [46]. Therefore, we believe that it forms an important research frontier of future study for the evolution of the electronic state in the iridate systems.

## 3. Conclusion

To summarize, this study demonstrates a pronounced electron doping effect into the Mott insulating iridate SL with a systematic manipulation of the valence state from $Ir^{4+}$ toward $Ir^{3+}$, which reveals the emergence of a novel metallic state at optimal doping level with suppressed magnetic states. Our studies not only make an important step forward for the long-sought electron doping in iridate Mott insulating systems, but also highlight the electric-field induced protonation as a promising pathway to manipulate phase diagram of complex oxides, which might lead to discovery of novel electronic states beyond standard chemical substitution and electrostatic gating approaches.

## 4. Experimental Section

*Thin film growth:* SrIrO$_3$/SrTiO$_3$ superlattices were fabricated with pulsed laser deposition method on SrTiO$_3$ (001) substrates, with an optimal condition of 700 °C, 0.1 mbar and 1.8 J/cm$^2$ (KrF, λ= 248 nm). The deposition process was *in-situ* monitored with a reflection high energy electron diffraction (RHEED) system, which guarantees a precise control of the atomic stacking sequence. Detailed structural characterizations can be found in *Ref .8*.

*Transport and XRD measurements:* The transport measurements were performed with a Quantum Design Dyna-Cool system. Hall bar structures were fabricated through lithography/etching method with a dimension of 1.6 mm × 0.4 mm and a thin layer of Pt was sputtered as electrode. *In-situ* measurements were performed with a slice of Pt as gate electrode, which was immerged together with the sample into as-received ionic liquid DEME-TFSI within a quartz bowl. To prevent electrochemical reaction between the ionic liquid and the electrode, an insulating separator (resin layer) was used to cover regions beyond the Hall bar. The presented *R-T* and in-plane magnetoresistance curves were obtained on a sample with thickness of 30 SL units and the presented Hall measurement was carried out on another sample with 25 SL units. X-ray diffraction (XRD) measurements were performed with a high-resolution diffractometer (Smartlab, Rigaku) using monochromatic Cu K$_{α1}$ (λ = 1.5406 Å) X-ray. The *In-situ* XRD measurements were performed on a SL sample with the size of 2 mm × 5 mm and thickness of 30 SL units, in which the amount of ionic liquid was carefully controlled to cover the entire samples and Pt electrode while keeping reasonable X-ray transmission.

*Operando Synchrotron X-ray absorption spectroscopy measurements:* *In-situ* X-ray absorption near edge spectroscopy (XANES) study during ionic liquid gating operation was carried out at the bending magnet beamline, 12-BM-B, at the Advanced Photon Source, Argonne National Laboratory. The linear polarized X-rays monochromatized by a pair of Si (111) single crystals with the energy resolution ~$1.4 \times 10^{-4}$ has a total flux of ~$2 \times 10^{11}$ photons/s onto the sample. The absorption spectra were collected by the fluorescence mode with samples mounted in a custom-designed X-ray electrochemical environmental cell integrated with *in-situ* electrochemical control of gating bias during the gating. A 13-element Ge drift detector (Canberra) was used to capture the X-ray fluorescence signals out of the thin film samples. A glancing incidence geometry (e.g., a few times of the substrate critical angle) was used to probe the signal contributed by the whole depth of the superlattice iridate films, as well as to reduce the elastic scattering background. A Ti metal foil and a Pt metal foil (Pt $L_3$ edge just nearby Ir $L_3$ edge) were used as the online calibration of the monochromator energy. The originally obtained XANES spectra were all normalized by fitting the pre-edge to zero and the post-edge to 1 using Ifeffit performed by the software Athena.

*First-principles calculation*: Spin-polarized density functional theory (DFT) calculations were performed using the projector augmented wave (PAW) method [47,48] with the PBEsol functional [49] as implemented in the Vienna Ab initio Simulation Package (VASP) code [50]. The energy cutoff of the plane-wave basis was set at 550 eV.

The valence states of Sr, Ir Ti, O, H were $4s^24p^65s^2$, $6s^15d^8$, $3p^64s^23d^2$, $2s^22p^4$, $1s^1$, respectively. In order to correct the on-site Coulomb interaction of the Ti 3*d* orbitals and Ir 5*d* orbitals, a rotationally invariant DFT+U method [51] was applied with U = 5.0 eV, J = 0.64 eV [52] and U = 2.0 eV, J = 0 eV [53] for Ti and Ir atoms, respectively. Spin-orbit coupling was also taken into accounts in our calculations. We relaxed all structures until the electronic convergence of $10^{-6}$ eV was reached and all coordinates were relaxed until the Hellmann-Feynman force on each atom was less than 0.01 eV/Å. The *k*-point meshes of 7 × 7 × 5 in the first Brillouin zone were used. The energy of inserting one proton is set as $E_H = -3.257$ eV; and the energy of extracting one oxygen from the superlattice unit to form a $H_2O$ molecule in ionic-liquid is set as $E_O = -7.78$ eV. Thus, the energy cost ($E_C$) to insert a proton in one superlattice unit is $E_C = (E_T - E_H) - E_0$; and the energy cost to induce one oxygen-vacancy from the superlattice unit is $E_C = (E_T + E_O) - E_0$. Here the $E_T$ is the total energy of each configuration with one proton intercalated or one oxygen extracted, and $E_0$ is the total energy of the pristine superlattice unit. The band structure calculations with the characteristics of total angular moment of $J_{eff}$ were performed by OPENMX code, which is based on the linear combination of pseudoatomic orbitals method [54].

## Acknowledgements

We thank Guangming Zhang, Fa Wang and Yi Zhou for fruitful discussions. This study was financially supported by the Basic Science Center Project of National Natural Science Foundation of China (NFSC) under grant No. 51788104, the NSFC under grant No. 51872155, the NSFC under grant No. 52025024, the Beijing Natural

Science Foundation (Grant No. Z200007), the Ministry of Science and Technology of China (2016YFA0301004), the Tsinghua University Initial Science Research Program (20203080003) and the Beijing Advanced Innovation Center for Future Chip (ICFC). This research used resources of the Advanced Photon Source, a U.S. Department of Energy (DOE) Office of Science User Facility operated for the DOE Office of Science by Argonne National Laboratory under Contract No. DE-AC02-06CH11357. Jian Liu acknowledges support from the National Science Foundation (Grant No. DMR-1848269) and the Office of Naval Research (Grant No. N00014-20-1-2809). Jia Li acknowledges support from the National Natural Science Foundation of China (11874036), Local Innovative and Research Teams Project of Guangdong Pearl River Talents Program (2017BT01N111) and Basic Research Project of Shenzhen, China (JCYJ20200109142816479). J.Y. would like to acknowledge funding from the State of Tennessee and Tennessee Higher Education Commission (THEC) through their support of the Center for Materials Processing.

**Author contributions**

M. W., L. H. and F. Y. contributed equally to this work. P. Y. and Jian. L. conceived this work. M. W. performed the transport, XRD, and SIMS measurements with helps from Y. W., J. Z., N. L. and S. S.. L. H. grew the thin films with helps from J. Y. and with the supervision of Jian. L. H. Z. and H. C. performed the synchrotron XAS measurements. F. Y., X. Y. and N. Z. performed the first-principles calculations under the supervision of Jia. L. M. W., L. H. and P. Y. wrote the manuscript. All authors discussed the results.

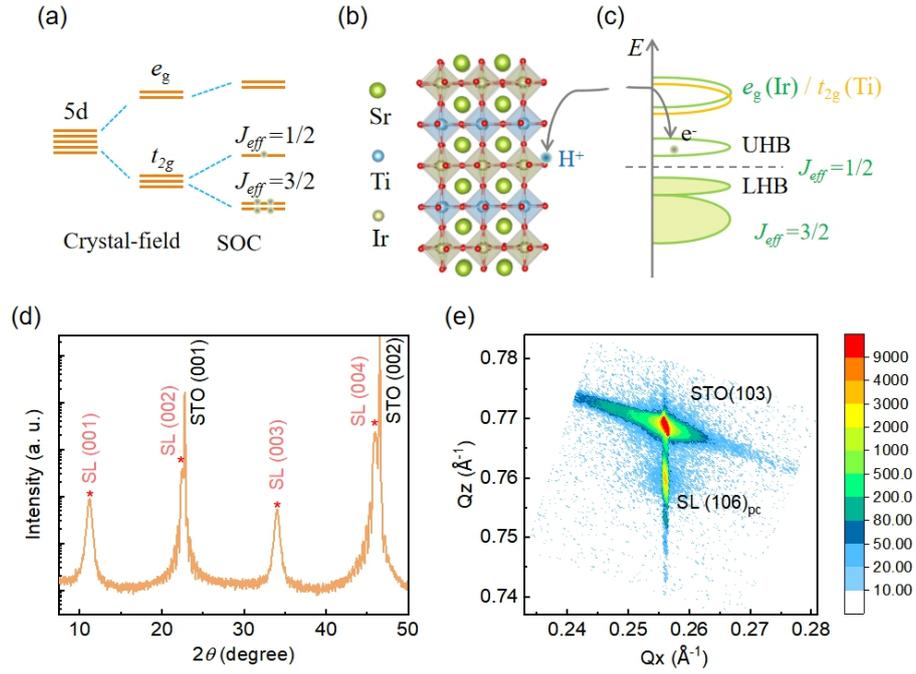

**Figure 1. $J_{eff}$ = 1/2 Mott insulating state in artificial constructed SrIrO$_3$/SrTiO$_3$ superlattice (SL)**. **a**) 5d orbital level splitting induced by the crystal-field splitting and strong spin-orbit coupling (SOC) within the IrO$_6$ octahedral framework. **b**) Illustration of an artificial constructed SrIrO$_3$/SrTiO$_3$ SL with the hydrogen ion (H$^+$, proton) intercalated into the lattice. Following the principle of charge neutrality, an electron is doped into the lattice along the protonation. **c**) Schematic illustration of band alignment near Fermi level in the SL according to the Mott-Hubbard model. The $J_{eff}$ = 1/2 level is split into an upper Hubbard band (UHB) and a lower Hubbard band (LHB), opening an electronic gap due to the electron correlation. Note that, the energy of Ti-$t_{2g}$ band locates much higher than the UHB and close to the Ir-$e_g$ band (*7*), which therefore would not be influenced by the electron filling. **d**) X-ray diffraction $2\theta$-$\omega$ scan of the SL grown on a SrTiO$_3$ (STO) substrate. **e**) Reciprocal space mapping (RSM) close to the pseudocubic (106) diffraction peak of the SL.

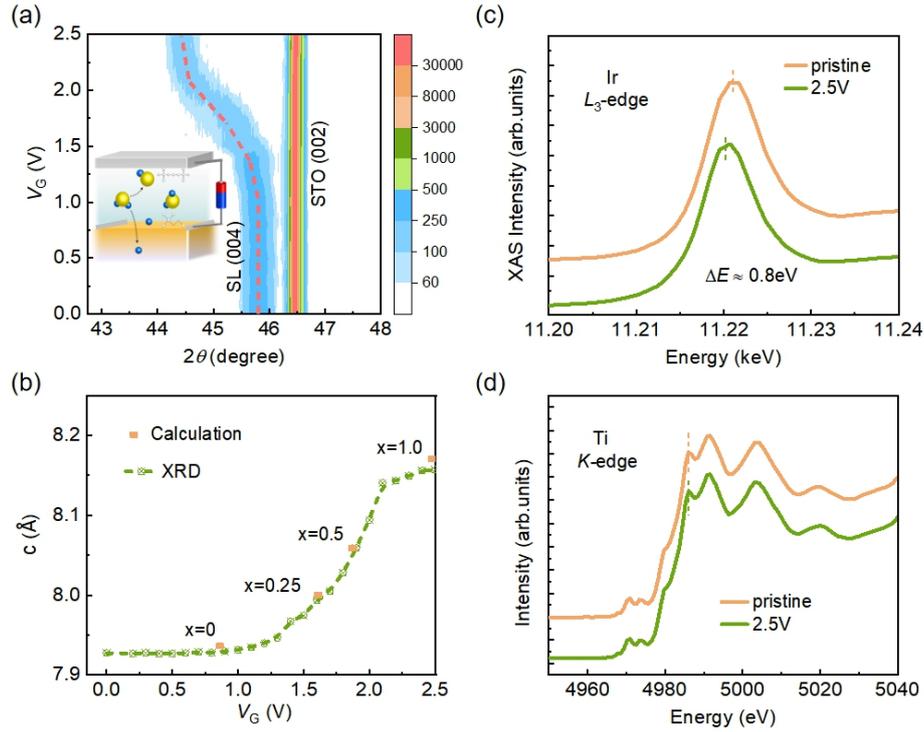

**Figure 2. Evolution of the structural and electronic states in SrIrO$_3$/SrTiO$_3$ SL upon protonation.** a) Evolution of the SL (004) diffraction peak along the increase of gate-voltage ($V_G$) during ionic-liquid gating. $2\theta$-$\omega$ scans were collected with an interval of 0.1 V. Insert shows an illustration of the experimental set-up during the ionic-liquid gating, in which the gate voltage induces an electrolysis of residual water molecular within ionic-liquid, and then drives the proton intercalation into the material under positive biased gating voltage. b) Evolution of *c* lattice constants obtained from *in-situ* XRD measurements at different gating voltages. Theoretically calculated lattice constants for H$_x$SrIrO$_3$/SrTiO$_3$ superlattice with different hydrogen contents are also shown for comparison. c, d) Direct comparison of *In-situ* hard X-ray absorption spectra for (c) Ir $L_3$-edge and (d) Ti $K$-edges between the pristine (orange) and gated state with $V_G$ = 2.5 V (olive). All spectra were *in-situ* measured with a small (grazing) incident angle to guarantee the XAS spectra obtained mainly from the SL, and a thicker (40 nm) sample was used during this measurement to enhance the signal.

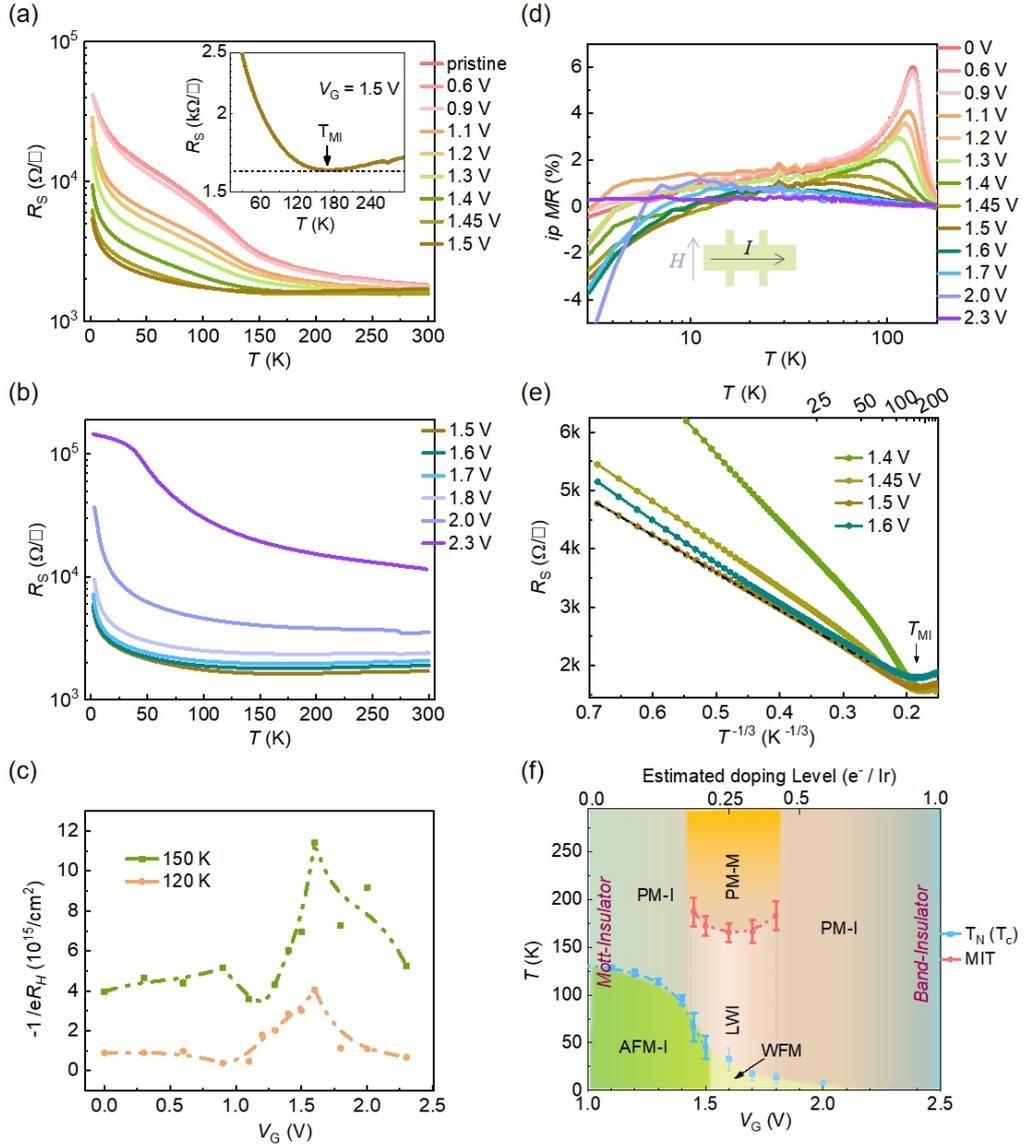

**Figure 3. *In-situ* transport measurements and phase diagram of SrIrO$_3$/SrTiO$_3$ SL through protonation induced electron filling. a, b)** Temperature dependent resistance of SL sample during ionic liquid gating with gate-voltage increasing from 0 V to 2.3 V. Inset is the zoomed-in version of the metal to weak-insulator transition with $V_G$ = 1.5 V. **c)** $V_G$ dependent Hall signal (-1/e$R_H$) at 120 K (orange) and 150 K (olive). **d)** Temperature dependent magnetoresistance under different $V_G$ with the magnetic-filed (9 T) applied along in-plane direction. Inset shows the illustration of the measurement configuration. **e)** Fitting of sheet-resistances at lower temperatures with $T^{-1/3}$ dependence. **f)** Sketched phase diagram of the proton intercalated SL. With the decreasing of temperature, the samples with lower electron doping ($V_G$ < 1.1 V)

show a pronounced paramagnetic insulator (PM-I) to antiferromagnetic insulator (AFM-I) transition. Samples at intermediate doping level (gated with $V_G$ around 1.45 V to 1.8 V) show a clear paramagnetic metal (PM-M) behavior at high temperatures, while form localized weak-insulating state (LWI) at lower temperatures. Samples with higher doping level (under $V_G > 2.3$ V) develop into a band-insulating-like state with vanished magneto response. At the same time, the Néel temperature ($T_N$, or $T_c$ for Curie Temperature) decreases with increasing doping level and a signature of weak-ferromagnetic order (WFM) emerges at $V_G > 1.5$ V. The electron doping levels were estimated according to the comparison of lattice constants (Figure 2b) obtained from *in-situ* XRD measurements (with different gating voltages) and theoretical calculated $H_xSrIrO_3/SrTiO_3$ with known electron doping levels.

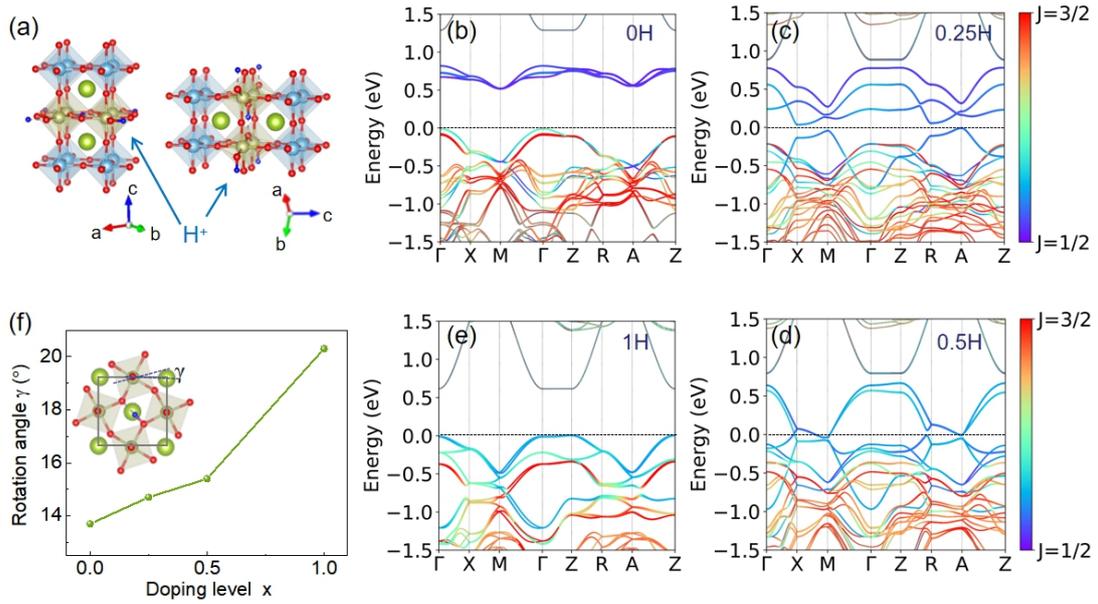

**Figure 4. Calculated crystal and electronic structures of SrIrO$_3$/SrTiO$_3$ SL through protonation. a)** Calculated crystalline structure of H$_{0.5}$SrIrO$_3$/SrTiO$_3$ SL with two proton (blue) intercalated into every four IrO$_6$ octahedral units. The intercalated protons bond with the oxygen ion at Ir-O planes, leading to electron doping into the Ir ion due to the principle of charge neutrality. **b-e)** Calculated band structures of (b) pristine SrIrO$_3$/SrTiO$_3$ SL, and proton intercalated SL with doping level of (c) 0.25, (d) 0.5 and (e) 1. The colored scale bars indicate the total angular momentum of $J_{eff}$ = 1/2 (blue) and $J_{eff}$ = 3/2 (red). **f)** Calculated IrO$_6$ octahedral rotation angle at different proton doping levels (x). Inset shows an illustration of the octahedral rotation of the H$_{0.5}$SrIrO$_3$ layer.